\documentclass[showpacs,prd,twocolumn]{revtex4}
\usepackage{graphicx}% Include figure files
\usepackage{dcolumn}% Align table columns on decimal point
\usepackage{bm}% bold math

\begin{document}

%%%%%%%%%%%%%%%%%%%%%%%%%%%%%%%%%%%%%%%%%%%%%%%%%%%%%%%%%%%%%%%%%%%%%%
%% Front Material %%%%%%%%%%%%%%%%%%%%%%%%%%%%%%%%%%%%%%%%%%%%%%%%%%%%
%%%%%%%%%%%%%%%%%%%%%%%%%%%%%%%%%%%%%%%%%%%%%%%%%%%%%%%%%%%%%%%%%%%%%%

\title{Stochastic Gravitational Wave Background from 
Cosmological Supernovae}

\author{Alessandra Buonanno$^{a,b}$, G{\"u}nter Sigl$^{a,b}$, 
Georg~G.~Raffelt$^c$, Hans-Thomas~Janka$^d$, Ewald~M\"uller$^d$} 

\affiliation{$^a$GReCO, Institut d'Astrophysique de Paris, C.N.R.S.,
98 bis boulevard Arago, F-75014 Paris, France\\
$^b$ AstroParticule et Cosmologie (APC),
11, place Marcelin Berthelot, F-75005 Paris, France\\
$^c$Max-Planck-Institut f\"ur Physik (Werner-Heisenberg-Institut),
F\"ohringer Ring~6, 80805 M\"unchen, Germany\\
$^d$Max-Planck-Institut f\"ur Astrophysik,
Karl-Schwarzschild-Str.~1, 85741 Garching, Germany}

\begin{abstract}
  Based on new developments in the understanding of supernovae (SNe)
  as gravitational-wave (GW) sources we estimate the GW background
  from all cosmic SNe.  For a broad range of frequencies around 1~Hz,
  this background is crudely comparable to the GW background expected
  from standard inflationary models.  While our estimate remains
  uncertain within several orders of magnitude, the SN GW background
  may become detectable by second-generation space-based
  interferometers such as the proposed Big-Bang Observatory (BBO).  By
  the same token, the SN GWs may become a foreground for searches of
  the inflationary GWs, in particular for sub-Hz frequencies where the
  SN background is Gaussian and where the BBO will be most sensitive.
  SN simulations lasting far beyond the usual cutoff of about 1~s are
  needed for more robust predictions in the sub-Hz frequency band.  An
  even larger GW background can arise from a hypothetical early
  population of massive stars, although their GW source strength as
  well as their abundance are currently poorly understood.
\end{abstract}


\maketitle

%%%%%%%%%%%%%%%%%%%%%%%%%%%%%%%%%%%%%%%%%%%%%%%%%%%%%%%%%%%%%%%%%%%%%%
%% Introduction %%%%%%%%%%%%%%%%%%%%%%%%%%%%%%%%%%%%%%%%%%%%%%%%%%%%%%
%%%%%%%%%%%%%%%%%%%%%%%%%%%%%%%%%%%%%%%%%%%%%%%%%%%%%%%%%%%%%%%%%%%%%%

\section{Introduction}

Core-collapse supernova (SN) explosions are among the most violent
astrophysical phenomena.  The total energy of about
$3\times10^{53}~{\rm erg}$ is primarily released in a neutrino burst
lasting for a few seconds. About 1\% goes into the actual explosion
while only a fraction of about $10^{-4}$ is emitted in visible light,
yet a SN can outshine its host galaxy.  In addition, SNe are expected
to be strong gravitational-wave (GW) sources. About 1~SN per second
takes place in the visible universe, but the neutrino burst has been
observed only once from SN~1987A, the closest SN in modern
history. GWs are even more difficult to detect and have never been
measured from any source.  On the other hand, the next galactic SN
probably will be observed with high statistics in several large
neutrino detectors as well as with the existing GW antennas. For the
first time there is a realistic opportunity to observe such a
cataclysmic event in all forms of radiation.

However, galactic SNe are rare so that the next SN neutrinos to be
observed could well be the diffuse neutrino background from all cosmic
SNe.  The limit from Super-Kamiokande~\cite{Malek:2002ns} already
touches the upper range of theoretical
predictions~\cite{Strigari:2003ig,Ando:2004hc,Strigari:2005hu}. The
proposed gadolinium upgrade~\cite{Beacom:2003nk} or a corresponding
future megatonne detector may well produce a positive detection.
Therefore, it is natural to ask the analogous question whether the
diffuse GW background from all past cosmic SNe will be important for
current or future GW antennas that search for signals from individual
astrophysical events or for the stochastic cosmic background that
probably arises in the very early universe during the inflationary
epoch.

The existing literature focuses on GWs from the SN bounce signal in
the kHz range that is relevant for bar detectors or ground-based laser
interferometers such as LIGO, VIRGO, GEO and TAMA. However, recent
studies of SNe as GW sources indicate that a much stronger signal may
be expected from the large-scale convective overturn that develops in
the delayed explosion scenario during the epoch of shock-wave
stagnation that may last for several 100~ms before the actual
explosion~\cite{Mueller:2003fs,Fryer:2004wi}.  Moreover, the relevant
frequencies reach below 1~Hz where they may be relevant for
space-based detectors. While we find that a first-generation
instrument such as LISA is not sufficient, the GW background from SNe
may become relevant for a second-generation detector such as the
proposed Big-Bang Observatory (BBO). 

While a detection of the GW background from SNe would be intriguing
from the perspective of SN physics, particularly in conjunction with a
future detection of the corresponding neutrino background, this
possibility could be a significant problem for searches of the inflationary
GW background that would probe a very early epoch of the universe. In
fact, if ordinary astrophysical foreground sources were seriously
expected to mask the inflationary GWs in some range of frequencies,
probably an instrument like the BBO should be designed to have its
optimal sensitivity in a different frequency range.  Therefore, a
reliable understanding of such foregrounds is an important task in
view of the long-term perspective of observing the big bang in GWs.

We begin our study in Sec.~II with a general expression for the GW
background from all cosmic SNe.  In Sec.~III we consider core-collapse
SNe as sources, whereas in Sec.~IV we turn to an early population of
massive stars. In Sec.~V we estimate the frequency range where the GW
background is Gaussian before concluding in Sec.~VI. We always use
natural units with $\hbar=c=1$.

%%%%%%%%%%%%%%%%%%%%%%%%%%%%%%%%%%%%%%%%%%%%%%%%%%%%%%%%%%%%%%%%%%%%%%
\section{Estimating the Gravitational Wave Background}
%%%%%%%%%%%%%%%%%%%%%%%%%%%%%%%%%%%%%%%%%%%%%%%%%%%%%%%%%%%%%%%%%%%%%%

We assume all SNe are identical GW sources defined by the Fourier
transform $\tilde{h}(f)\equiv\int_{-\infty}^{+\infty}dt\,e^{-i2\pi
ft}h(t)$ of the dimensionless strain amplitude $h(t)$, which is
proportional to deviations from spherical symmetry of the
energy-momentum tensor.  In general, $h(t)$ is obtained from numerical
simulations. However, the contribution of asymmetric neutrino emission
is explicitly~\cite{Epstein:1978dv,mueller2}
\begin{equation}
\label{Lnu}
h(t)=\frac{2G_{\rm N}}{D}\int_{-\infty}^{t-D} 
dt^\prime\,L_\nu(t^\prime)q(t^\prime)\,,\label{memory}
\end{equation}
where $G_{\rm N}$ is Newton's constant, $D$ the distance to our
``standard SN,'' and $L_\nu(t)$ the neutrino luminosity. Further,
$q(t)\leq1$ is an asymmetry parameter defined as the angle-dependent
neutrino luminosity folded over an angular function; for details see
Eqs.~(27)--(29) in Ref.~\cite{mueller2}.

Equation~(\ref{memory}) shows that $h(t)$ converges to a constant
value $h_\infty$ for $t\gtrsim t_e$, with $t_e$ the neutrino emission
time scale of a few seconds.  Therefore, the neutrino burst causes
$h(t)$ to jump from zero to a non-vanishing constant
value, an effect called the ``burst with memory''~\cite{Smarr:1977fy, 
Epstein:1978dv, Turner:1978jj, gw_memory,Burrows:1995bb}. 
We now write
\begin{eqnarray}
2\pi f \, \tilde{h}(f) &=&-i\int_{-\infty}^{+\infty} dt
\, e^{-i 2 \pi f t}\,\dot{h}(t)
\nonumber \\
&\simeq& -i\frac{2G_{\rm N}}{D}\,\int_{-\infty}^{+\infty} dt
\,L_\nu(t)q(t)\,,
\end{eqnarray}
where in the second step we used Eq.~(\ref{Lnu}) and assumed
$f \ll (2 \pi t_e)^{-1}$, which is known as the zero-frequency  
limit~\cite{Smarr:1977fy, Epstein:1978dv, Turner:1978jj,bontz_price}.
This implies
\begin{eqnarray}
f|\tilde{h}(f)|&\simeq &\frac{|h_\infty|}{2\pi}=
\frac{G_{\rm N}}{\pi D}
\left\langle q\right\rangle E_\nu\label{low_freq} \\
&\simeq&
2.6\times10^{-19}\,\left\langle q\right\rangle
\left(\frac{10\,{\rm kpc}}{D}\right)
\left(\frac{E_\nu}{3\times10^{53}\,{\rm erg}}\right)\,, 
\nonumber
\end{eqnarray}
where $E_\nu=\int_{-\infty}^{+\infty} dt^\prime\, L_\nu(t^\prime)$ is
the total emitted neutrino energy and $\left\langle
q\right\rangle\equiv \int_{-\infty}^{+\infty} dt^\prime\,
L_\nu(t^\prime)q(t^\prime)/E_\nu$ is the luminosity-weighted average
neutrino anisotropy.  Although the zero-frequency limit
Eq.~(\ref{low_freq}) is valid for $f \ll (2 \pi t_e)^{-1}$, we will
use it to evaluate the GW signal at $f \lesssim (2 \pi t_{\rm
sim})^{-1}$, where $t_{\rm sim} < t_e$ is the maximal simulation time
(less than a second), and then we continuously extend $|\tilde{h}(f)|$
to lower frequencies. Current simulations do not cover the strain
spectrum below fractions of a Hertz.

The energy density in GWs at frequency $f$ per logarithmic frequency
interval in units of the cosmic critical density $\rho_{\rm c} =
3H_0^2/(8 \pi G_{\rm N})$ can be written as~\cite{Phinney:2001di}
\begin{equation}
\label{los}
\Omega_{\rm gw}(f)=\frac{16\pi^2D^2}{15G_{\rm N}\rho_{\rm c}} 
\int_0^\infty dz\,
\frac{R_{\rm SN}(z)}{1+z}\left|\frac{dt}{dz}\right|
f_z^3\left|\tilde{h}(f_z)\right|^2\,,
\end{equation}
where $R_{\rm SN}(z)$ is the SN rate per comoving volume,
$f_z\equiv(1+z)f$, and $\tilde{h}(f)$ is the strain spectrum of an
individual SN at distance $D$.  The cosmological model enters with
$|dt/dz|=[(1+z)H(z)]^{-1}$ and, for a flat geometry,
\begin{equation}
\label{cosmo}
H(z)= H_0\left[\Omega_{\rm M}(1+z)^3+\Omega_{\Lambda}\right]^{1/2}\,.
\end{equation}
We will use the parameters $\Omega_{\rm M}=0.3$,
$\Omega_{\Lambda}=0.7$, and $H_0=h_0\,100~{\rm km}~{\rm s}^{-1}~{\rm
Mpc}^{-1}$ with $h_0=0.72$.

The SN rate as seen from Earth is
\begin{equation}\label{duty1}
\int_{0}^{\infty}dz\,\frac{R_{\rm SN}(z)}{1+z}\frac{dV}{dz}=
\int_{0}^{\infty}dz\,R_{\rm SN}(z)\frac{4\pi r^2(z)}{(1+z)H(z)}\,,
\end{equation}
where $dV/dz$ is the fractional volume element, the cosmic expansion
rate at redshift $z$ is given by Eq.~(\ref{cosmo}), and $r(z)$ is the
comoving coordinate, $dr=(1+z)dt$.

%%%%%%%%%%%%%%%%%%%%%%%%%%%%%%%%%%%%%%%%%%%%%%%%%%%%%%%%%%%%%%%%%%%%%%
\section{Supernovae}
%%%%%%%%%%%%%%%%%%%%%%%%%%%%%%%%%%%%%%%%%%%%%%%%%%%%%%%%%%%%%%%%%%%%%%

The cosmic star-formation rate and, as a consequence, the
core-collapse SN rate is reasonably well known at redshifts $z\lesssim
5$~\cite{Daigne:2004ga}, but becomes rather uncertain at larger $z$.
The SN rate per comoving volume is often expressed as
\begin{equation}
\label{evol}
R_{\rm SN}(z)=R_{\rm SN}^0\times\left\{\begin{array}{ll}
(1+z)^\beta & \mbox{for $z<1$}\\
2^{\beta-\alpha}(1+z)^\alpha &
\mbox{for $1\leq z\leq20$}\end{array}\right..
\end{equation}
If not otherwise stated, we will use $\beta=2.7$, $\alpha=0$, and the
present-day rate $R_{\rm SN}^0=2\times 10^{-4}\,{\rm Mpc}^{-3}\, {\rm
yr}^{-1}$. These values are consistent with the Super-K limits on the
diffuse neutrino background~\cite{Malek:2002ns}. The parameter
$0\lesssim\alpha\lesssim2$ is much less constrained than $\beta$ and
$R_{\rm SN}^0$ because $\alpha$ influences the rate only for $z>1$
where the neutrinos are redshifted below the Super-K
threshold~\cite{Strigari:2003ig,Ando:2004hc,Strigari:2005hu}.  This
situation may change if low-threshold detectors such as the gadolinium
upgrade of Super-K~\cite{Beacom:2003nk} become operative.

\begin{figure}[t]
\includegraphics[width=0.48\textwidth,clip=true]{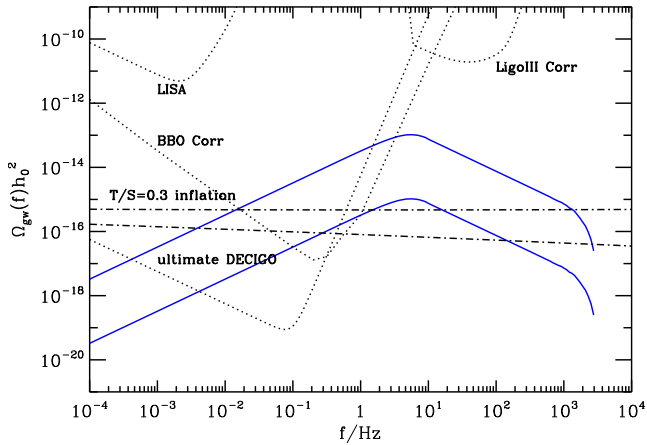}
\caption{Spectrum of GW background from Eq.~(\ref{los}) for a
simulation from Ref.~\cite{Fryer:2004wi}.  Solid lines: the spectrum
is continuously extended with the low-$f$ tail according to
Eq.~(\ref{low_freq}). The upper curve is obtained from Fig.~9 of
Ref.~\cite{Fryer:2004wi}, whereas the lower is the same shifted
downwards by a factor 100, which is considered a more realistic
estimate.  Horizontal lines: GW stochastic spectrum produced during
slow-roll inflation and evaluated from Eq.~(6) of
Ref.~\cite{Turner:1996ck}. We assume $T/S=0.3$ for the ratio of the
tensorial and scalar contributions to the cosmic microwave background
radiation (CMBR) anisotropy and consider two values $\pm 10^{-3}$ of
the running of the tensorial power-law index. Dotted lines:
sensitivities of the indicated detectors.  The BBO sensitivity is
approximate and may change slightly in the final design.}
\label{fig1}
\end{figure}

For the strain spectrum $\tilde{h}(f)$ of an individual event we first
use the recent 3-dimensional asymmetric simulation shown in Fig.~9 of
Ref.~\cite{Fryer:2004wi}.  We plot in Fig.~\ref{fig1} the resulting GW
background together with an estimate scaled downward by a factor of
100.  The latter GW background with a total energy released in GWs of
about $8\times10^{-11}\,M_\odot$ should be considered more
realistic~\cite{Fryer:2004wi}.  We also show sensitivities of LISA,
BBO~\cite{bbo}, LIGO (or EGO) correlated third
generation~\cite{AB:2002}, and the ultimate DECIGO~\cite{Seto:2001qf},
which would be a quantum limited space-based interferometer with
mirror masses of 100~kg. For some frequencies, our lower estimate for
the GW background from SNe can be comparable to the most optimistic GW
background from slow-roll inflation (horizontal lines).

Next, we use the three models s15r, s11nr180, and pns180 from recent
two-dimensional (2D) core-collapse simulations based on a detailed
implementation of spectral neutrino transport or following the
long-time evolution of the newly born neutron star with good
resolution, respectively~\cite{Mueller:2003fs}.  Models s15r and
s11nr180 are fully 2D, i.e.\ axially symmetric, stellar core collapse
and SN simulations of a rotating 15$\,M_\odot$ and a non-rotating
11.2$\,M_\odot$ star with very strong core convection after shock
formation. Convection takes place both in the neutrino-heated
postshock layer and inside the nascent neutron
star~\cite{Buras:2003,Janka:2004jb}.  The simulations were done with a
full spectral treatment of neutrino transport and neutrino-matter
interactions with gravitational redshift effects taken into
account. An approximation to 2D transport was used which takes into
account the dependence of the transport and neutrino-matter
interactions on the laterally varying conditions in the star. It is
based on the solution of the coupled set of radiation moments
equations and Boltzmann transport equation, assuming locally radial
neutrino fluxes.  Moreover, order $v/c$ velocity terms in the moments
equations due to the motion of the stellar fluid and the momentum
transfer to the stellar plasma by neutrino pressure gradients were
taken into account (``ray-by-ray plus'' approximation; for details see
Ref.~\cite{Buras:2005}). Model pns180 is a 2D hydrodynamic simulation
of a non-rotating, convecting ``naked'' proto-neutron star, which was
evolved through its neutrino-cooling phase from shortly after core
bounce until about 1.2~s later
\cite{Keil:1996,Keil-PhD,Janka:1997id,Janka2001}.  The SN explosion
was assumed to have taken off and therefore accretion and postshock
convection were not included in this model. Neutrino transport was
treated by flux-limited equilibrium diffusion, again assuming that the
neutrino flux follows locally radial gradients according to the
varying conditions in the 2D stellar background. Assuming neutrinos in
equilibrium with the stellar medium also implied that effects of
neutrino advection and ``radiation compression'' and the fluid
acceleration by neutrino pressure gradients were accounted for. In all
models the neutrino emission developed a time-dependent anisotropy
because of the two-dimensionality of the evolving stellar
background. The lateral variation leads to a non-vanishing value of
$q(t)$. The approximations in the description of 2D transport tend to
yield an underestimation of the power in emission asymmetries with
large angular scales and low frequency, but an overestimation of
high-frequency and short-wavelength anisotropies in the neutrino
emission. The reason for this behavior is that local fluctuations in
the convective layers around the neutrinosphere translate by the
mostly radial transport into small-scale fluctuations of the neutrino
emission. In reality, however, multi-dimensional transport is much
more non-local and the emitted radiation therefore contains averaged
information from all parts of the emitting surface, thus averaging
over short-wavelength variability of the hydrodynamic medium of the
star (for a discussion in the context of neutrino transport in SN
cores see Ref.~\cite{Walder:2005}).

\begin{figure}
\includegraphics[width=0.48\textwidth,clip=true]{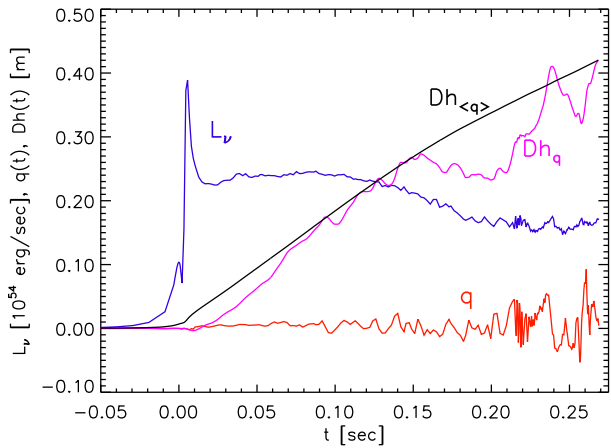}
\caption[...]{Neutrino luminosity $L_\nu(t)$, an\-isotropy $q(t)$, and
GW strain $h(t)$ times distance from anisotropic neutrino emission
only, see Eq.~(\ref{Lnu}), for model s15r of
Ref.~\cite{Mueller:2003fs}, as functions of time after bounce.  We
also show the GW strain $h(t)$ times distance from anisotropic
neutrino emission, using the average anisotropy $\langle q \rangle =
0.45 \% $.}
\label{fig2}
\end{figure}

\begin{figure}
\includegraphics[width=0.48\textwidth,clip=true]{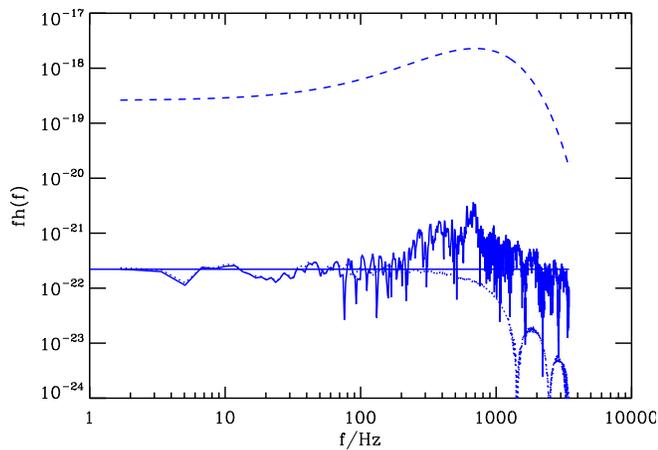}
\caption[...]{GW source spectra: Solid and dotted ragged lines are
total and neutrino contribution, respectively, for model s15r of
Ref.~\cite{Mueller:2003fs} at distance $D=10$~kpc.  The straight solid
line is the low-$f$ tail Eq.~(\ref{low_freq}) with $h_\infty$ from
Eq.~(\ref{memory}), using $q(t)$ and $L_\nu (t)$ of Fig.~\ref{fig2},
leading to $f|\tilde{h}(f)|\simeq2.21\times10^{-22}$.  The dashed line
is the schematic spectrum for PopIII stars of Eq.~(\ref{popIII}).}
\label{fig3}
\end{figure}

We first focus specifically on model s15r which gives more optimistic
GW signals. Figures~\ref{fig2} and~\ref{fig3} show neutrino luminosity
$L_\nu(t)$, anisotropy $q(t)$, and GW strain $h(t)$ and GW source
spectra $f|\tilde{h}(f)|$, respectively. Figure~\ref{fig3} shows that
anisotropic neutrino emission dominates at low frequencies.  Note that
the simulation stops at about 250~ms after the bounce when only about
$0.5\times10^{53}$~erg has been emitted in neutrinos, i.e.\ about 1/6
of the total, and when $\int dt\,L_\nu(t)q(t)=2.55\times10^{50}$~erg.
The average anisotropy over this time period is $\langle
q\rangle\simeq0.45\%$, but from Fig.~\ref{fig2} we can see that
$|q(t)|$ tends to increase. If $q(t)$ were constant, according to
Eq.~(\ref{memory}), $h(t)$ would increase proportional to the emitted
neutrino energy, whereas for a fluctuating $q(t)$ we could expect
that $h(t)$ increases roughly as the square root of that energy.
Therefore, an estimate of the GW signal $h(t)$ when $3
\times10^{53}$~erg has been emitted in neutrinos, can be obtained
introducing in the strain an enhancement factor between $\sim\sqrt{6}$
and $\sim6$.

\begin{figure}
\includegraphics[width=0.48\textwidth,clip=true]{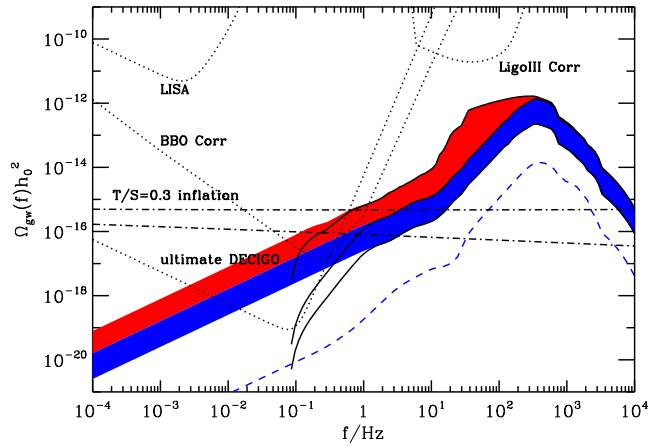}
\caption[...]{GW background for model s15r with rotation (colored
bands and solid lines) and model s11nr180 without rotation (dashed
line) from Ref.~\cite{Mueller:2003fs}.  The source spectra have been
continuously extended using the zero-$f$ tail, Eq.~(\ref{low_freq}),
for $f\lesssim1$~Hz, except for the solid lines. The blue band and
lower solid lines for model s15r reflect the plausible range $6$--$36$
of the enhancement factor to correct for the limited time of the
simulation. For model s11nr180 the lower enhancement factor of 6 was
assumed. We always use $\alpha=0$ in Eq.~(\ref{evol}), except for the
red band and upper solid line which show the difference between
$\alpha=0$ and $\alpha=2$ (for enhancement factor 36) for model s15r.}
\label{fig4}
\end{figure}

The resulting GW backgrounds are shown in Fig.~\ref{fig4}. For model
s15r the blue band corresponds to the uncertainty of the enhancement
factor discussed above, and the red band is a rough estimate of the
uncertainty due to redshift evolution. For $100~{\rm Hz}\lesssim
f\lesssim1$~kHz the signal is dominated by the convective motion in
the neutrino-heated postshock layer, whereas at lower frequencies it
is mostly due to asymmetric neutrino emission. This scenario
corresponds to a total energy release in GWs of
$\simeq1.8\times10^{-8}\,M_\odot$ for the most conservative
enhancement factor. Note that, although the zero-frequency signal is
lower, the GW energy release is higher than in the case of
Fig.~\ref{fig1} because it is dominated by high frequencies.

Equations~(\ref{low_freq}) and~(\ref{los}) show that for our standard
redshift evolution ($\alpha = 0$ and $\beta = 2.7$) and
$E_\nu\simeq3\times10^{53}$~erg liberated in neutrinos, to completely
mask the inflation background down to 10~mHz we would need an average
anisotropy $\langle q\rangle\simeq 6$\%, which is much higher than
what is predicted by our most optimistic simulation model s15r where
$\langle q \rangle \simeq 0.45 \%$.  We note that average quadrupole
anisotropies $\langle q \rangle$ of the order of 1\% are currently not
excluded by astrophysical observations. In fact a 1\% dipole
anisotropy corresponds to observed kick velocities $\simeq400\,{\rm
km}\,{\rm s}^{-1}$ for a neutron star of mass $\simeq1.4M_\odot$, but
we do not know whether and how quadrupole and dipole anisotropies are
correlated.

For $\alpha\lesssim2$ in Eq.~(\ref{evol}), the integral in
Eq.~(\ref{los}) converges rather quickly.  Cutting the integral at
$z=5$ instead of $z=20$, for example, lowers the predictions by less
than a factor~2. In contrast, the poorly constrained redshift
evolution for $z>1$ causes more significant uncertainties as
illustrated in Fig.~\ref{fig4}.

\begin{figure}
\includegraphics[height=0.48\textwidth,clip=true,angle=90]{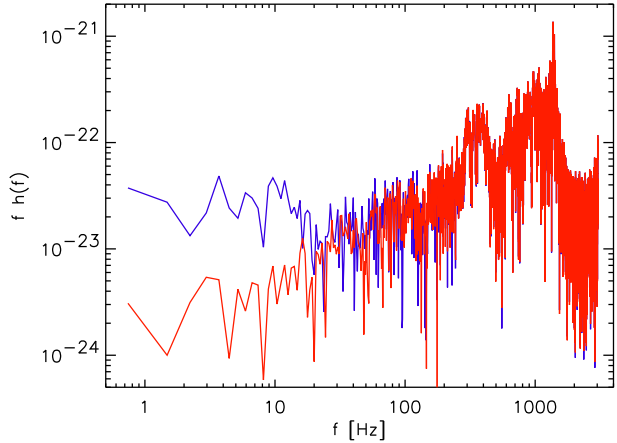}
\caption[...]{GW source spectra from mass motions plus an\-iso\-tropic
neutrino emission for the proto-neutron star model pns180 of
Ref.~\cite{Mueller:2003fs} at a distance of 10~kpc. The blue curve is
for the full simulation with the fluctuating $q(t)$ integrated over
1.2~s, whereas for the red curve $q(t)$ was replaced with
$\left\langle q\right\rangle\sim3\times10^{-5}$.}
\label{fig5}
\end{figure}

To test the low-frequency dependence on the fluctuations of the
neutrino anisotropy, we analyze GW signals for the proto-neutron star
model of Ref.~\cite{Mueller:2003fs} whose evolution was followed for
1.2~s.  In Fig.~\ref{fig5} we compare the GW spectra for (i) the full
simulation integrated over 1.2~s and (ii)~the spectrum obtained from
replacing the time-dependent anisotropy $q(t)$ of the neutrino
emission with its luminosity-weighted average over the 1.2~s duration
of the simulation, $\langle q\rangle\sim3\times10^{-5}$.  The huge
difference of $\langle q\rangle$ for models s15r and pns180 arises
from the rotation in model s15r, that leads to a global, slowly
growing deformation and thus produces a systematic trend on $q(t)$. In
contrast, the merely convective fluctuations in model pns180 on
time scales of a few milliseconds cause a much lower time-averaged
value of~$q$.

\begin{figure}
\includegraphics[width=0.48\textwidth,clip=true]{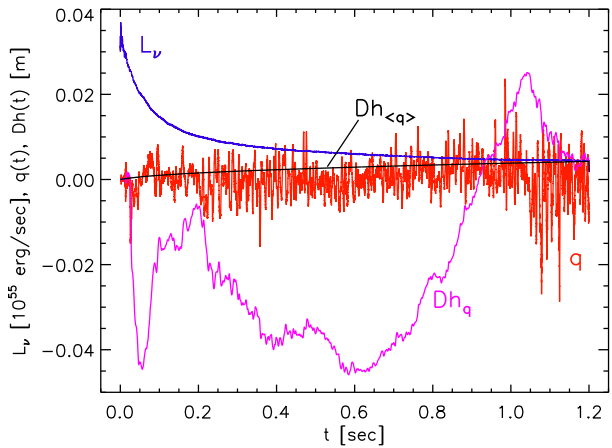}
\caption[...]{Neutrino luminosity $L_\nu(t)$, an\-isotropy $q(t)$, and
GW strain $h(t)$ times distance from anisotropic neutrino emission
only, see Eq.~(\ref{Lnu}), for model pns180 of
Ref.~\cite{Mueller:2003fs}, as functions of time after bounce.  We
also show the GW strain $h(t)$ times distance from anisotropic
neutrino emission, using the average anisotropy $\langle q \rangle = 3
\times 10^{-5}$.}
\label{fig8}
\end{figure}

From Fig.~\ref{fig5} we conclude the following.  For $f\gtrsim30\,$Hz,
replacing $q(t)$ by $\left\langle q\right\rangle$ makes no significant
difference (the blue curve is hidden below the red one) because the
signal is dominated by convection and thus does not depend on $q(t)$.
In contrast, for $1\,{\rm Hz}\lesssim f\lesssim10\,$Hz, fluctuations
in $q(t)$ increase the amplitude by a factor $\simeq10$. We also note
the following technical point due to the properties of Fourier
transformations and the SN simulations used here: In the proto-neutron
star model pns180 the frequencies covered by the simulation do not
extend to the regime where the zero-frequency limit applies, because
$L_\nu(t)q(t)$ oscillates with an almost constant amplitude over the
entire duration of the simulation as seen from Fig.~\ref{fig8}, and
thus has structure on time scales over which $\exp(i2\pi ft)$ still
oscillates even in the lowest frequency bin.  If neutrino emission
would cease right after the end of the pns180 simulation,
$f\tilde{h}(f)$ obtained with the fluctuating $q(t)$, i.e.\ the blue
curve in Fig.~\ref{fig5}, should approach
$f|\tilde{h}(f)|\to4\times10^{-24}$ for $f\ll1\,$Hz. In contrast, in
model s15r the product $L_\nu(t)q(t)$ starts out very small over more
than half of the simulation time and tends to increase at later times,
as seen from Fig.~\ref{fig2}.  Therefore, in the lowest frequency bin
covered by the simulation of model s15r, $\exp(i2\pi ft)$ does not
fluctuate significantly over time scales on which $L_\nu(t)q(t)$ has
structure, and the zero-frequency limit overlaps with the lowest
frequencies covered by the simulation.

\begin{figure}
\includegraphics[width=0.48\textwidth,clip=true]{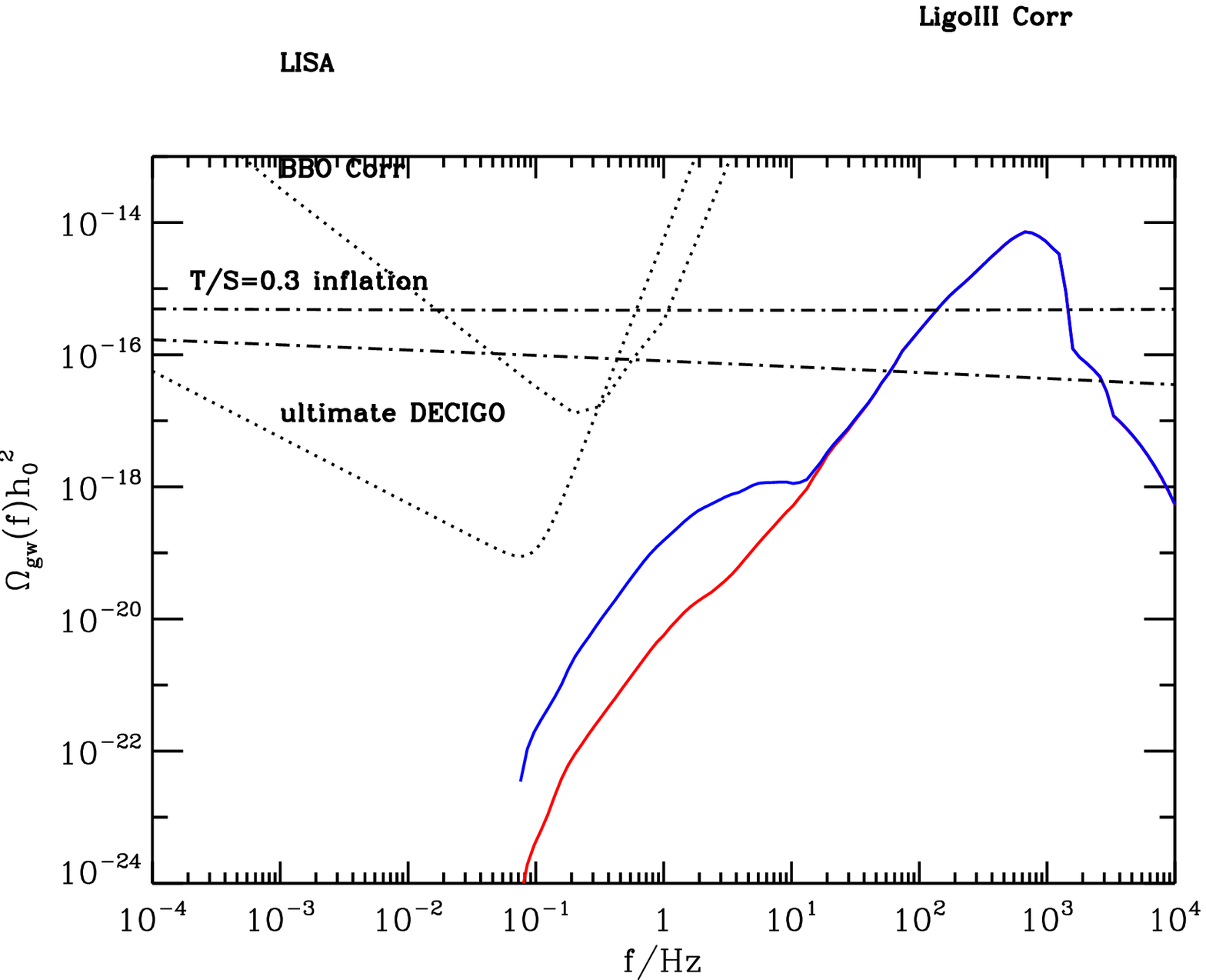}
\caption[...]{GW background resulting from the source spectrum for the
proto-neutron star model pns180 of Ref.~\cite{Mueller:2003fs} shown in
Fig.~\ref{fig5} without continuation to $f\lesssim1\,$Hz. We use
$\alpha=0$ in Eq.~(\ref{evol}) and an enhancement factor of 3 (about
1/3 of the neutrino energy is emitted until simulation end). The solid
blue line thus corresponds to the blue line in Fig.~\ref{fig5}
(fluctuating $q(t)$), whereas the solid red line is for $q(t)$
replaced by $\langle q \rangle$ (red line in Fig.~\ref{fig5}). Other
notation as in Fig.~\ref{fig4}.}
\label{fig6}
\end{figure}

The GW background for the proto-neutron star models of Fig.~\ref{fig5}
is shown in Fig.~\ref{fig6}. Due to the redshift effect those GW
spectra lie in part in the frequency range of the BBO space-based
detectors. Here we are only interested in the relative amplitude
between cases (i) and (ii), and not in their absolute amplitude, which
is well below the sensitivity of space-based detectors.
   
Finally, the ``tail'' of the neutrino luminosity during neutrino
cooling (times later than 1--2~s after neutron star birth) is known
from spherically symmetric neutron star cooling
simulations~\cite{Keil:1995,Pons:1998mm,Thompson:2001ys} and can be
roughly approximated by $L\propto t^{-n}$ with $n$ around unity (we
use $n = 1.1$). This in principle could be used to extrapolate the
signal to low frequencies. However, the evolution of the anisotropy
parameter $q(t)$ during the late phases of proto-neutron star cooling
($\gtrsim\,$1--2~s after bounce) has not been determined by models
yet. Any assumption about its behavior after $\sim\,$1$\,$s implies
major uncertainties.  As discussed above, a constant anisotropy
parameter or a fluctuating one can give very different predictions for
the GW spectrum in different core-collapse models: For $1\,{\rm
Hz}\lesssim f\lesssim10\,$Hz, the GW source amplitude is suppressed by
roughly a factor 10 in model pns180 if $q(t)$ is replaced by a
constant equal to its luminosity-weighted average (see
Fig.~\ref{fig5}). In contrast, for model s15r, the GW source amplitude
is almost identical in these two cases in that same frequency range,
recall Fig.~\ref{fig3}. Similarly, if we evaluate the GW source
spectrum by extrapolating the neutrino luminosity decline by $L\propto
t^{-1.1}$ and replacing a fluctuating $q(t)$ with a constant equal to
its luminosity-weighted average, we would obtain an $h(f)$ smaller by
one order of magnitude in the frequency range of space-based
detectors.  Note that uncertainties in the GW source amplitude of the
order of 10 translate into uncertainties in $\Omega_{\rm gw}$ of a
factor $\sim100$.
 
In conclusion, temporal fluctuations of the neutrino anisotropy are
extremely important for predicting the stochastic GW signal at
frequencies $f\sim0.1$~Hz so that longer simulations are needed to
make robust predictions.

%%%%%%%%%%%%%%%%%%%%%%%%%%%%%%%%%%%%%%%%%%%%%%%%%%%%%%%%%%%%%%%%%%%%%%
\section{Population III stars}
%%%%%%%%%%%%%%%%%%%%%%%%%%%%%%%%%%%%%%%%%%%%%%%%%%%%%%%%%%%%%%%%%%%%%%

Core collapse events of the hypothetical population~III (PopIII)
generation of first stars~\cite{Bromm:2003vv,Ripamonti:2005ri} could
be much more efficient emitters of GWs than today's SN
populations~\cite{Fryer:2000my}.  Lacking more detailed numerical
models, we assume a schematic standard case of a mass of
$300\,M_\odot$, with a shape of the GW spectrum at $D=10$~kpc
resembling that for ordinary SNe shown in Fig.~\ref{fig3},
\begin{equation}
f|\tilde{h}(f)|=2.6\times10^{-19} \left(1+\frac{f}
{0.2~{\rm kHz}}\right)^3 e^{-f/0.3~{\rm kHz}}\,.
\label{popIII}
\end{equation}
The normalization implies 
\begin{equation}
E_{\rm gw}=\frac{16\pi^2 D^2}{15G_{\rm N}}\,\int df\,|f\tilde{h}(f)|^2
\simeq2 \times 10^{-3}M_\odot
\end{equation}
for the total energy emitted in GWs.  This very rough estimate is
based on hydrodynamical mass motions in 2D
geometry~\cite{Fryer:2000my}.  If we use the zero-frequency limit of
Eq.~(\ref{low_freq}) and $E_\nu \simeq10^{55}$~erg emitted in
neutrinos as given in Fig.~9 of Ref.~\cite{Fryer:2000my} for the model
with rotation of their Table 1, we find that the average anisotropy
would correspond to $\langle q\rangle \simeq 3\%$ in this case. While
this is significantly larger than the values expected for ordinary
SNe, it may not be unrealistic, given that PopIII stars tend to rotate
more rapidly and explode more violently.

Finally, we assume that the PopIII rate is concentrated around a
redshift $z_{\rm III}=15$, which may explain
reionization~\cite{Kogut:2003et}.  We thus write for the PopIII rate
$R_{\rm SN}(z)=R\delta(z-z_{\rm III})$, where $R$ is a normalization
constant. We further assume that all PopIII stars have a progenitor
mass $M_{\rm III}=300M_\odot$. The total number of PopIII events per
comoving volume that occurred up to $z=0$ can then be written as
\begin{equation}
  \int_0^\infty dt R_{\rm SN}(z)
  =\frac{R}{(1+z_{\rm III})H(z_{\rm III})}
  =f_{\rm III}\,n_\gamma\,\eta\,\frac{m_N}{M_{\rm III}}\,,
\end{equation}
where $n_\gamma\simeq410\,{\rm cm}^{-3}$ is the CMBR photon density at
redshift zero, $\eta\simeq6.3\times10^{-10}$ is the number of baryons
per CMBR photon, $m_N$ is the nucleon mass, and $f_{\rm III}$ is the
fraction of all baryons going through PopIII stars. Eliminating $R$,
it then follows that the rate observed at Earth, $R_{\rm III}$ given
in Eq.~(\ref{duty1}), is directly proportional~to~$f_{\rm III}$,
\begin{equation}
  R_{\rm III}\simeq4\pi r^2(z_{\rm III})\,f_{\rm III}n_\gamma\eta
  \,\frac{m_N}{M_{\rm III}}\,.\label{f_III}
\end{equation}
For the parameters given above this yields
$R_{\rm III}\simeq0.2\,(f_{\rm III}/10^{-3})\,{\rm s}^{-1}$.

\begin{figure}[t]
\includegraphics[width=0.48\textwidth,clip=true]{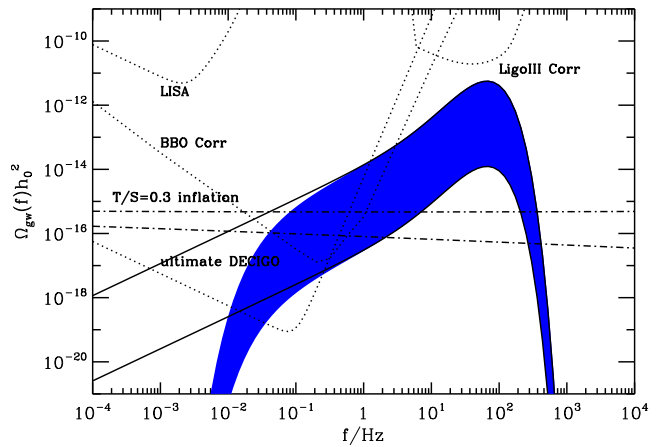}
\caption[...]{GW background for our standard PopIII core collapse case
and event rates $4.4\times10^{-4}~{\rm s}^{-1}<R_{\rm III}<0.2~{\rm
s}^{-1}$. For the blue shaded band the source spectrum shown in
Fig.~\ref{fig3} was cut off at $f\le1\,$Hz and in this sense does not
rely on the zero-frequency limit. The solid lines are obtained from
continuing the source signal by the zero-frequency limit.}
\label{fig7}
\end{figure}

The resulting GW background is shown in Fig.~\ref{fig7} as a band
delimited by two extreme assumptions about the total cosmic PopIII
core-collapse rate observed from Earth, i.e.\ $R_{\rm III}\simeq
0.2~{\rm s}^{-1}$~\footnote{More specifically,
Ref.~\cite{Daigne:2004ga} finds a rate $R_{\rm III}\simeq0.7~{\rm
s}^{-1}$ for PopIII stars with mass $270 \mbox{--} 500 M_\odot$ and
$R_{\rm III}\simeq0.013~{\rm s}^{-1}$ for PopoIII stars with mass $140
\mbox{--} 260 M_\odot$. We notice that an event rate as high as
$\simeq0.7~{\rm s}^{-1}$ is still compatible with metal production
since it is evaluated for massive non-rotating stars which collapse
directly into black holes, thus they do not produce metals.} (upper
edge of band) and $\simeq4.4\times10^{-4}~{\rm s}^{-1}$
\cite{Wise:2004zp} (lower edge).  The zero-frequency limit would
affect the GW spectrum only at frequencies below 0.05~Hz.  We remark
that certain models attempting to explain the near infrared excess
with PopIII stars require baryon fractions as high as $f_{\rm III}
\simeq 0.05$~\cite{Madau:2005wv}, corresponding to rates $R_{\rm
III}\simeq10\,{\rm s}^{-1}$ and GW backgrounds a factor $\sim 50$
still higher than the upper band shown in Fig.~\ref{fig7}. Rates
corresponding to $R_{\rm III}\simeq1\,{\rm s}^{-1}$ are discussed in
Ref.~\cite{Mesinger:2005du}.  The problem with such high-rate
scenarios is to avoid overproduction of metals.

Whereas neutrinos from PopIII stars are redshifted to energies
probably too low to be detected~\cite{Iocco:2004wd}, the GW background
for $10~{\rm mHz}\lesssim f\lesssim1~{\rm kHz}$ could be dominated by
these objects.

%%%%%%%%%%%%%%%%%%%%%%%%%%%%%%%%%%%%%%%%%%%%%%%%%%%%%%%%%%%%%%%%%%%%%%
\section{Gaussianity}
%%%%%%%%%%%%%%%%%%%%%%%%%%%%%%%%%%%%%%%%%%%%%%%%%%%%%%%%%%%%%%%%%%%%%%

Finally, we estimate the duty cycle of the background at a given
frequency $f$. It is defined as the product of the rate of events and
the time scale of interest, $f^{-1}$.

The duty cycle is roughly given by dividing the total cosmic SN rate,
Eq.~(\ref{duty1}), by the frequency. A probably more realistic
estimate is obtained by folding the rate integral over redshift with
the redshift dependent contribution to the GW signal given by the
integrand in Eq.~(\ref{los}). For model s15r of
Ref.~\cite{Mueller:2003fs} and the evolution discussed around
Eq.~(\ref{evol}) these two estimates give approximately $33\,(f/{\rm
Hz})^{-1}$ and $4\,(f/{\rm Hz})^{-1}$, respectively.

The signal from ordinary SNe becomes Gaussian for $f\lesssim1$~Hz, in
agreement with a cosmic SN rate of about $1~{\rm s}^{-1}$.  In
contrast, the backgrounds in the $10^2 \mbox{--}10^3$ Hz range are not
Gaussian, consistent with Ref.~\cite{Schneider:1999us} where only the
bounce signal was considered that lasts about $10^{-3}$~s. At these
frequencies the duty factor is $\lesssim1\%$. The putative PopIII
background becomes Gaussian only for $f\lesssim R_{\rm III}$, i.e.\
probably for $f\lesssim0.1$~Hz.

%%%%%%%%%%%%%%%%%%%%%%%%%%%%%%%%%%%%%%%%%%%%%%%%%%%%%%%%%%%%%%%%%%%%%%
\section{Conclusions}                          \label{sec:conclusions}
%%%%%%%%%%%%%%%%%%%%%%%%%%%%%%%%%%%%%%%%%%%%%%%%%%%%%%%%%%%%%%%%%%%%%%

The GW background from cosmological core-collapse SNe at frequencies
below 1~Hz is Gaussian. Its power is uncertain by several orders of
magnitude, mostly due to uncertainties of SNe as GW sources. The most
important uncertain parameters are the asymmetry parameter $\langle q
\rangle$ of late-time neutrino emission and the parameter $\alpha$
determining the star-formation rate at redshifts $z>1$ [see
Eq.~(\ref{evol})].  The most optimistic current simulations 
(for stellar core collapse with rotation) predict
$\langle q \rangle \sim 0.45 \%$, whereas Super-K limits on the
diffuse neutrino background constrain $\alpha$ in the range
$0\lesssim\alpha\lesssim2$.  Using these parameter values and
extrapolating the GW signal with the zero-frequency limit,
Eq.~(\ref{low_freq}), to the frequency band of space-based detectors,
we showed that the GW background from SNe could be comparable to the
maximum signal expected from standard inflationary models and thus
could be detected by second-generation space-based detectors.
However, to completely mask the inflationary background down to
10~mHz, we would need a very large asymmetry of $\langle q \rangle
\sim 6 \%$, that is not compatible with current simulations. By using
the proto-neutron star model of Ref.~\cite{Mueller:2003fs}, for which
the simulation continues until 1.2~s, we observed that the GW spectra
at frequencies of interest for future space-based detectors depends
strongly on the fluctuations of $q(t)$. Predictions could differ by a
few orders of magnitude if $q(t)$ continues to fluctuate for all or
part of the neutrino cooling phase even after 1~s of neutron star
cooling.

Numerical simulations and stronger constraints or detections of the
corresponding neutrino background will lead to an improved prediction
of the GW background.  However, accurate calculations of the GW
emission from core collapse are hampered by our incomplete knowledge
of important input physics like the equation of state at nuclear and
supernuclear densities, of initial conditions like the structure of
stellar cores at the onset of gravitational collapse, and the amount
of rotation and size of magnetic fields in the pre-collapse core. More
reliable GW predictions also require a better understanding of the
explosion mechanism and improvements in the numerical modeling of
SNe. Quantitatively meaningful calculations will have to be done in
3D, will have to include a decent treatment of general relativistic
effects, and will have to be done with full 3D neutrino transport in
order to determine the neutrino emission anisotropy on all angular
scales and at all frequencies. Three-dimensional effects are expected
to reduce the emission anisotropy of neutrinos because of smaller
convective structures \cite{mueller2} and because of the tendency of
multi-dimensional transport to reduce the power on high spatial
frequencies. In contrast, significant rotation and large-scale
magnetic fields might cause global emission anisotropies with possibly
constant anisotropy parameter and high power at low frequencies. It is
left to future generations of SN models to shed light on these
important questions.

PopIII stars could give a particularly strong contribution, masking
almost completely the inflationary background, if we use the upper
range~\cite{Daigne:2004ga} of the very uncertain rate
predictions~\cite{Wise:2004zp,Madau:2005wv,Mesinger:2005du} in the
literature and assume a total energy emitted in GWs of $2 \times
10^{-3} M_\odot$. For typical event rates and/or lower values of the
total GW energy emitted, the GW spectrum from PopIII stars could be
comparable to or lower than the one expected from inflation.  In this
case the zero-frequency limit would influence the GW spectrum only at
frequencies below 0.05~Hz. In any event, the largest problem for a
reliable prediction of the GW background from stellar core collapse is
the almost complete lack of theoretical understanding of massive
PopIII stars as GW sources and the extreme uncertainty of their
abundance, in fact the uncertainty of their very existence.

In summary, the uncertainties of the stochastic GW background from the
core collapse of ordinary SNe and from PopIII stars is large, but it
is intriguing that its power is in the ball park of the maximum GW
background expected from inflationary models. Of course, the
inflationary GW background itself is also very uncertain and could be
smaller than the benchmark shown in our figures.  If stochastic GWs
are detected with a future instrument like the BBO, it may be
challenging to disentangle the inflationary GWs from those caused by
SNe or the collapse of PopIII stars.

%%%%%%%%%%%%%%%%%%%%%%%%%%%%%%%%%%%%%%%%%%%%%%%%%%%%%%%%%%%%%%%%%%%%%%
%% Acknowledgments %%%%%%%%%%%%%%%%%%%%%%%%%%%%%%%%%%%%%%%%%%%%%%%%%%%
%%%%%%%%%%%%%%%%%%%%%%%%%%%%%%%%%%%%%%%%%%%%%%%%%%%%%%%%%%%%%%%%%%%%%%

\acknowledgments

We thank Tom Abel, Luc Blan\-chet, Fr\'ed\'eric Daigne, Daniel Holz,
Scott Hughes, Sterl Phinney, and Joe Silk for informative
discussions. In Garching and Munich, this work was partly supported by
the Deut\-sche For\-schungs\-ge\-mein\-schaft under grants SFB-375
``Astro Particle Physics'' and SFB-Transregio~7 ``Gravitational Wave
Astronomy''. The models of the Garching group were calculated on the
IBM p690 {\em Regatta} of the Rechenzentrum Garching.

%%%%%%%%%%%%%%%%%%%%%%%%%%%%%%%%%%%%%%%%%%%%%%%%%%%%%%%%%%%%%%%%%%%%%%
%% Bibliography %%%%%%%%%%%%%%%%%%%%%%%%%%%%%%%%%%%%%%%%%%%%%%%%%%%%%%
%%%%%%%%%%%%%%%%%%%%%%%%%%%%%%%%%%%%%%%%%%%%%%%%%%%%%%%%%%%%%%%%%%%%%%

%\clearpage

\end{document}